\documentclass[journal,transmag]{IEEEtran}

\usepackage{subfigure}
\usepackage{cite}

\ifCLASSINFOpdf
   \usepackage[pdftex]{graphicx}
   \graphicspath{{.}}
   \DeclareGraphicsExtensions{.pdf,.jpeg,.png}
\else
  \usepackage[dvips]{graphicx}
  \graphicspath{{.}}
  \DeclareGraphicsExtensions{.eps}
\fi

\usepackage{epstopdf}

\usepackage[cmex10]{amsmath}
\usepackage{array}

\ifCLASSOPTIONcompsoc
  \usepackage[caption=false,font=normalsize,labelfont=sf,textfont=sf]{subfig}
\else
  \usepackage[caption=false,font=footnotesize]{subfig}
\fi
 \usepackage{dblfloatfix}
\begin{document}
%
\title{Magnetovolume Effects in Heusler Compounds via First-Principles Calculations}


\author{\IEEEauthorblockN{J. N. Gon\c{c}alves\IEEEauthorrefmark{1},
J. S. Amaral\IEEEauthorrefmark{1,2}, and
V. S. Amaral\IEEEauthorrefmark{1}}
\IEEEauthorblockA{\IEEEauthorrefmark{1}Departamento de F\'isica and CICECO, Universidade de Aveiro, 3810-093 Portugal}
\IEEEauthorblockA{\IEEEauthorrefmark{2}IFIMUP and IN-Institute of Nanoscience and Nanotechnology, Rua do Campo Alegre, 678, 4169-007 Porto, Portugal}%
\thanks{Corresponding author: J.~N. Gon\c{c}alves (email: joaonsg@ua.pt).}}

\markboth{CU - Ab-initio and transport 2}%
{Shell \MakeLowercase{\textit{et al.}}: Bare Demo of IEEEtran.cls for Journals}
%



\IEEEtitleabstractindextext{%
\begin{abstract}
Heusler alloys are promising for several applications, including magnetic refrigeration,  due to high magnetocaloric and magnetovolume effects. One way to optimize this potential is by increasing the magnetovolume effect. Using density functional theory with the Korringa-Kohn-Rostoker method, we calculate the effective exchange interaction energies and corresponding mean field Curie temperature as a function of the volume (hydrostatic pressure) in several L2$_1$-type Co$_2$YZ Heusler alloys. Different qualitative trends and signs of the pressure derivatives of the Curie temperature and moments are found among these compounds, discussed and compared with previous calculations and experiments.  
\end{abstract}

}

\maketitle

\IEEEdisplaynontitleabstractindextext

%
\IEEEpeerreviewmaketitle

\section{Introduction}
%
%
%
%
\IEEEPARstart{T}{he} magnetocaloric effect (MCE) is a common property to all magnetic materials, and is the basis of magnetic refrigeration, an eco-friendly and efficient technology~\cite{Tishin03}. The desired magnetic properties of a room-temperature magnetic refrigerant material are a Curie temperature ($T_C$) near room-temperature, and high maximum temperature derivative of magnetization ($dM/dT$) in a wide temperature range. Other performance aspects come into play, in particular thermal properties such as specific heat and thermal conductivity, precursor availability (such as rare-earths) and the time/cost of synthesis methods. In order to increase the $dM/dT$ of a given magnetic refrigerant, one may aim to increase the saturation magnetization value $M_s$, but magnetovolume coupling can play an even more important role. Indeed, if this coupling is sufficiently strong, a material can present a first-order magnetovolume phase transition and a giant MCE~\cite{Pecharsky97}.

Some transition-metal based ferromagnetic Heusler alloys show a strong MCE, particularly those that present a martensitic phase transition, such as Ni-Mn-(Ga,Sn,In) based alloys~\cite{Krenke05}. For these systems, and for Mn based Heusler alloys in general, several experimental studies have investigated the pressure dependence of magnetic properties (e.g.\ \cite{Kanomata12} and references therein). First-principles calculations of the pressure dependence of Curie temperature in Ni$_2$MnSn have also been done~\cite{Sasioglu05}, where a study up to high pressures found a non-monotonic dependence, which was explained by the competing influences of the electron hopping and magnetic moment changes with lattice parameter. Alloys based on this system doped with metals were also studied by first-principles~\cite{Bose11}. However, in this respect, the cobalt based systems (Co$_2$YZ) are less explored. These systems are also interesting for spintronics, since they are among the Heusler alloys with the highest spin polarization~\cite{landolt-bornstein}. In this work, we explore several compositions of Heusler ferromagnets via first-principles calculations, with general formula Co$_2$YZ. We assess their magnetovolume coupling ($J$ and $T_C$ dependence with lattice volume), and  $dM/dP$. 

\section{Technical Details}
The \textit{ab initio} density functional calculations use the Korringa-Kohn-Rostoker Green's function method as implemented in the \textsc{spr-kkr} code~\cite{SPR-KKR}, with no shape approximations (full-potential method). Spin-polarization is considered, with the scalar-relativistic approximation. The exchange-correlation functional uses the local-spin-density approximation (LSDA), with the Vosko-Wilk-Nusair parameterization~\cite{Vosko80}. The maximum angular momentum cutoff of the $l$-expansion is taken as $l=3$. The reciprocal space is sampled with a $22\times22\times22$ $\mathbf{k}$-mesh.

The exchange energies of an effective Heisenberg Hamiltonian 
are then calculated for the different compounds and volumes, with the magnetic force theorem, approach of Liechtenstein et.\ al~\cite{Liechtenstein87}, in spherical clusters around the different atoms, of radius $3a$ ($a$ is the lattice constant). The Curie temperature is then estimated using these energies with the mean field approximation. It has been shown that for Co$_2$FeSi and NiMn$_2$Sb only the first few neighbors contribute with significant interactions, and our Curie temperatures are likely converged as a function of this cutoff radius for all compounds~\cite{Thoene09,Sasioglu05}. In a four multi-sublattice system such as the Heusler alloys, this calculation is done by using the coupled set of equations
\begin{equation}
\frac{3}{2}k_BT_C^{\text{MFA}}\langle s_\mu\rangle = \sum_\nu J_0^{\mu\nu}\langle s_\nu\rangle,
\end{equation}
where $\mu$, $\nu$ are different sublattices, $J_0^{\mu\nu}=\sum_{\mathbf{r}\neq 0}J_{0\mathbf{r}}^{\mu\nu}$ and $\langle s_\nu\rangle$ is the average $z$ component of the unit vector $\mathbf{s}_\mathbf{r}^\nu$ in the direction of the magnetic moment at site ($\nu$,$\mathbf{r}$).
This is equivalent to solving the eigenvalue problem 
\begin{equation}
(\mathbf{\Theta}-T\mathbf{I})\mathbf{S}=0
\end{equation}
where $\mathbf{I}$ is the identity matrix, $\mathbf{S}$ is the vector of $\langle s_\nu\rangle$, $\Theta_{\mu\nu}=(2/3k_B)J_0^{\mu\nu}$ and the largest eigenvalue of $\mathbf{\Theta}$ gives the Curie temperature~\cite{Sasioglu05,Anderson63}.

\section{Results}
\begin{figure}[ht!]
\centering
\includegraphics[width=2.5in]{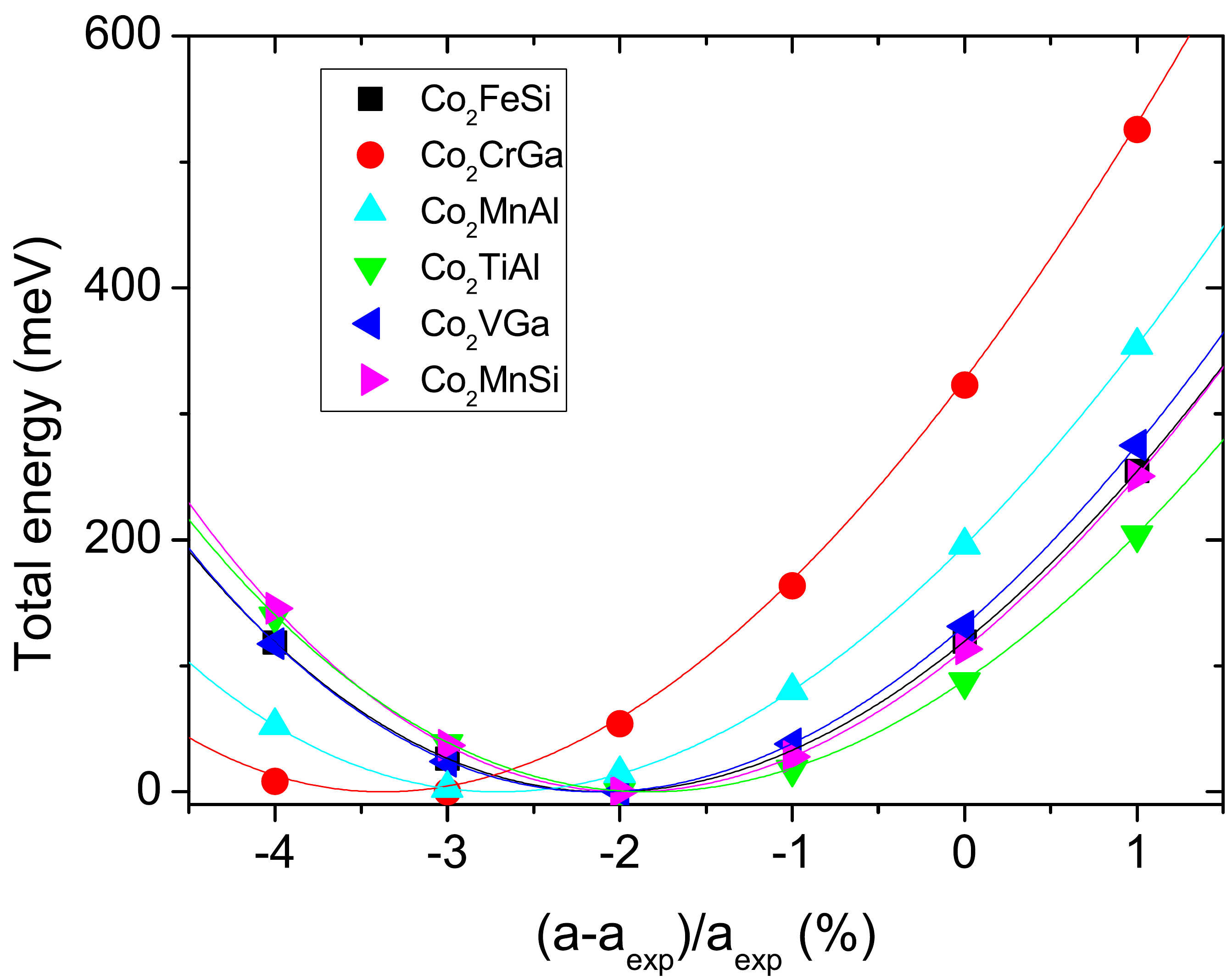}
\caption{Total energy as a function of the lattice parameter. The lines are obtained by the fits to the Birch-Murnaghan equation of state. The zero of energy corresponds to the lowest value of each fit.}
\label{fig_energy}
\end{figure}

The calculations are performed on Heusler ferromagnetic compounds with the L2$_1$ structure type, of the general formula Co$_2$XY, namely Co$_2$FeSi, Co$_2$CrGa, Co$_2$MnAl, Co$_2$TiAl, Co$_2$VGa, and Co$_2$MnSi. There have been studies of these compounds e.g.\ for the exchange interactions and finite temperature properties (Thoene et al.~\cite{Thoene09}), but the magnetovolume effects were not examined. For each of these cases, calculations with variation of the lattice constant are performed, with respect to the experimental values, from $-4\%$ to $+1\%$ in steps of $1\%$. The obtained total energies, shown in Fig.~\ref{fig_energy}, were fitted to a Birch-Murnaghan~\cite{Birch47} equation of state. The zero of energy in Fig.~\ref{fig_energy} is the obtained minimum in each case. See also the table for the theoretical lattice parameters and the bulk moduli. The minimum of energy is around $-2$ to $-3.5\%$, at slightly lower values than experiment, as expected for the LSDA. The bulk modulus, at least for Co$_2$FeSi, is in good agreement (experimental value is $240$\ GPa~\cite{Garg11}). 

\begin{figure}[ht!]
\centering
\includegraphics[width=2.7in]{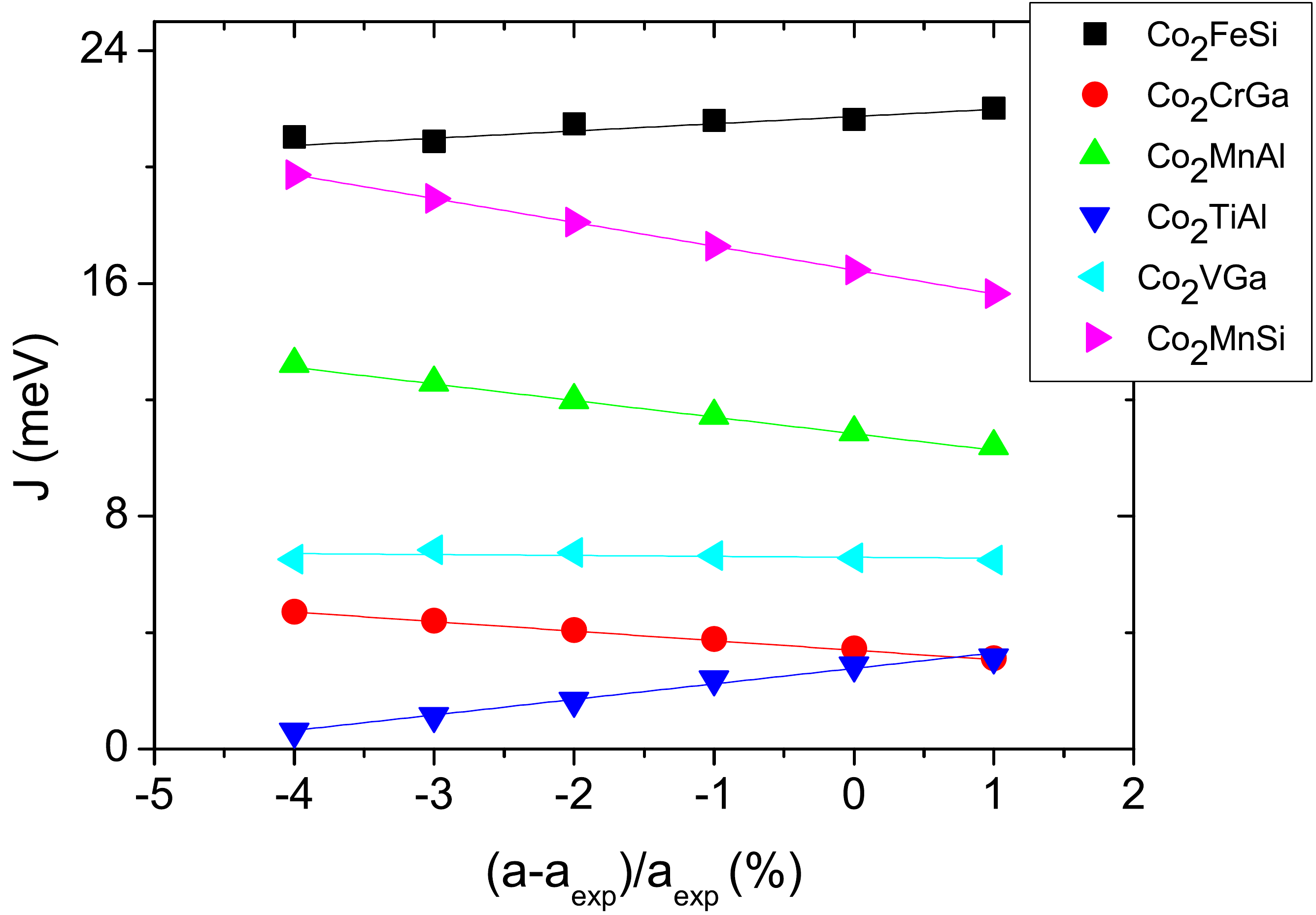}
\caption{Largest pairwise exchange interaction in the compounds studied with variation of the lattice parameter. The lines are linear fits to the data.}
\label{Dominant_J}
\end{figure}

 The dominant pairwise exchange interactions in each compound (between Co and the other transition metal in Co$_2$FeSi, Co$_2$CrGa , Co$_2$MnAl, and Co$_2$MnSi, between Co1 and Co2 in Co$_2$TiAl, and Co$_2$VGa) are presented in Fig.~\ref{Dominant_J}, with its volume dependence. These results are indicators of the compounds with largest magnetovolume effects. 

\begin{figure}[ht!]
\centering
\includegraphics[width=2.7in]{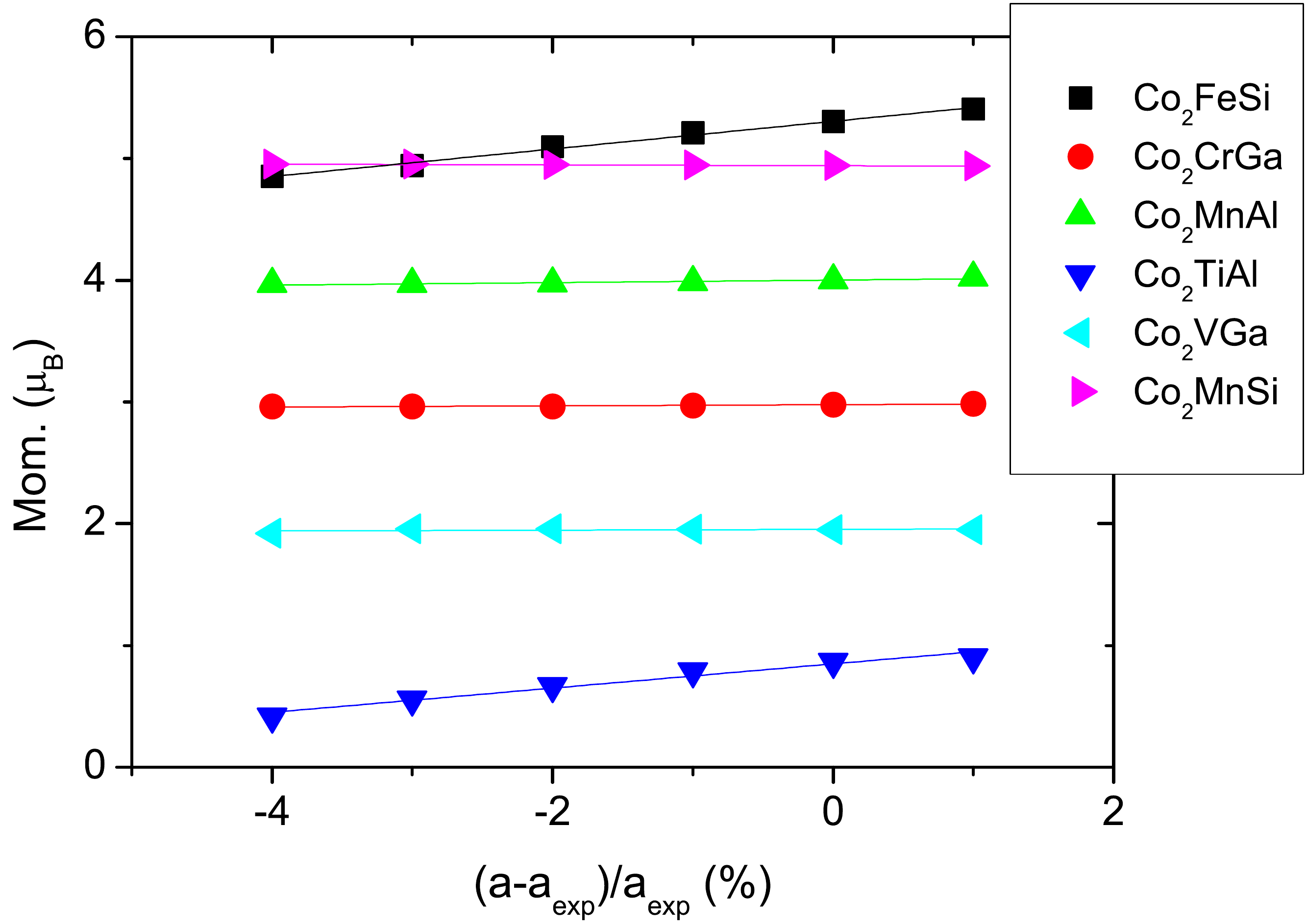}
\caption{Total spin moment as a function of the lattice parameter. The lines are linear fits to the data.}
\label{fig_moments}
\end{figure}

The total spin moment and its volume variation is also shown in Fig.~\ref{fig_moments}. The results are close to an integer value in four of the cases, with negligible volume variation, indicating half-metallic character for its electronic structure. For the other two cases, Co$_2$FeSi and Co$_2$TiAl, there is significant variation with volume, the spin moments increasing with lattice parameter, presumably due to the decreased overlap of atomic states at higher volumes. The moment of Co$_2$FeSi is underestimated with respect to the high experimental value of $6\ \mu_B$~\cite{Wurmehl06}, but is in agreement with the previous calculations~\cite{Thoene09}. Thoene et al. tried to improve the result with the LSDA+U approximation, obtaining an increase of the magnetic moment to $5.82\ \mu_B$, but the Curie temperature also increased $300$-$500$\ K, in disagreement with experiment. Therefore, in this case better DFT approximations and Curie temperature calculation approaches should be used to produce both moments and Curie temperatures in agreement with experiment.

For the case of Co$_2$VGa, the variation of magnetic moments and Curie temperature with pressure was studied experimentally and with complementary density functional calculations of the magnetic moment using the projector augmented-wave method~\cite{Kanomata10}. We found a moment close to $2$\ $\mu_B$, as predicted by the Slater-Pauling rule, with no significant changes with volume, except a slight decrease at the lowest lattice parameter ($5.55$\ \AA{}). This is consistent with the other calculation~\cite{Kanomata10} in the same volume range. That calculation was done up to lower volumes, showing a more pronounced spin moment decrease.

\begin{figure}[ht!]
\centering
\includegraphics[width=2.5in]{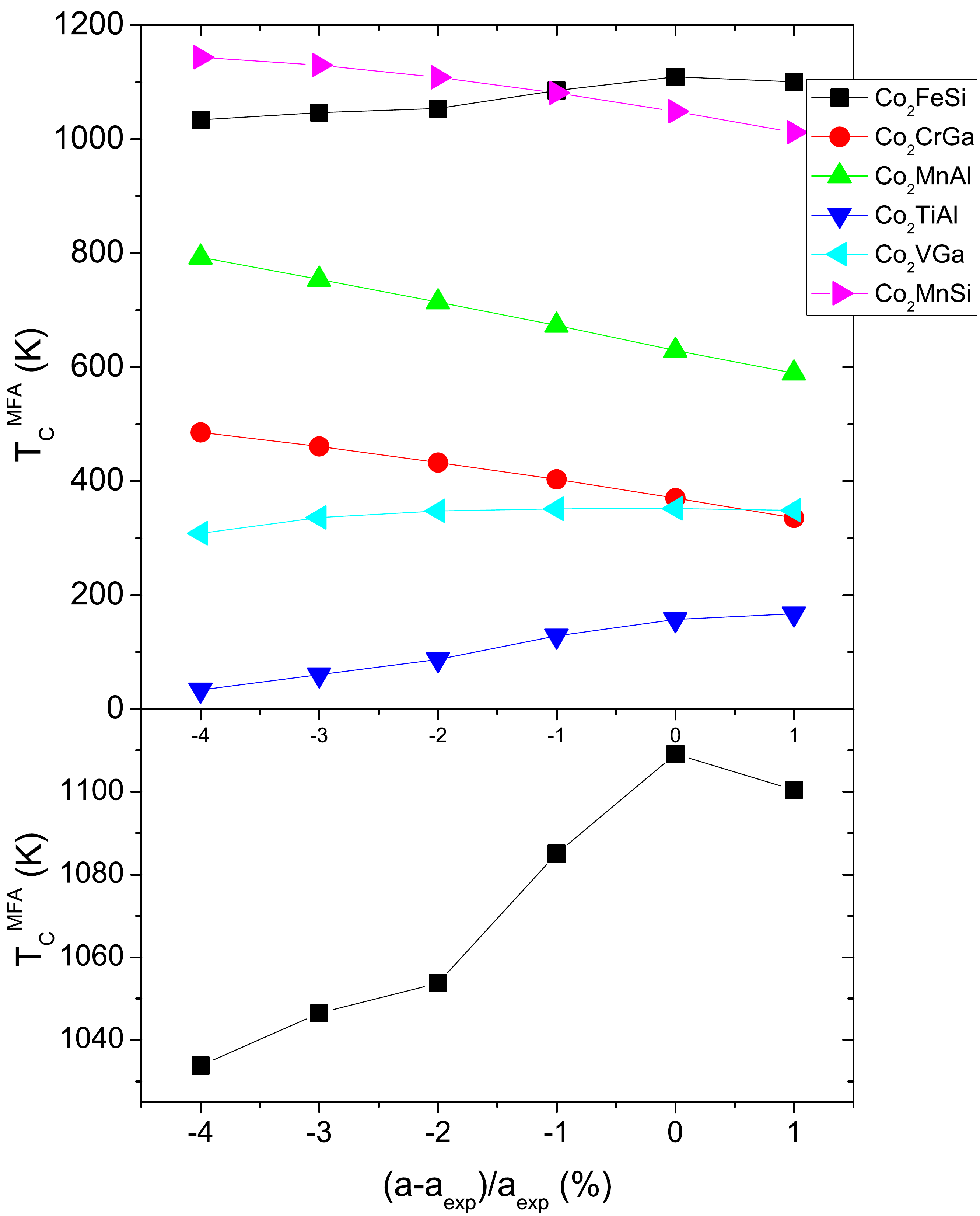}
\caption{Top: Lattice parameter dependence of the calculated Curie temperatures for all the Heusler compounds studied. The lines are guides to the eye. Bottom: Detail of results for Co$_2$FeSi.}
\label{fig_TCs}
\end{figure}

\begin{figure}[ht!]
\centering
\includegraphics[width=2.5in]{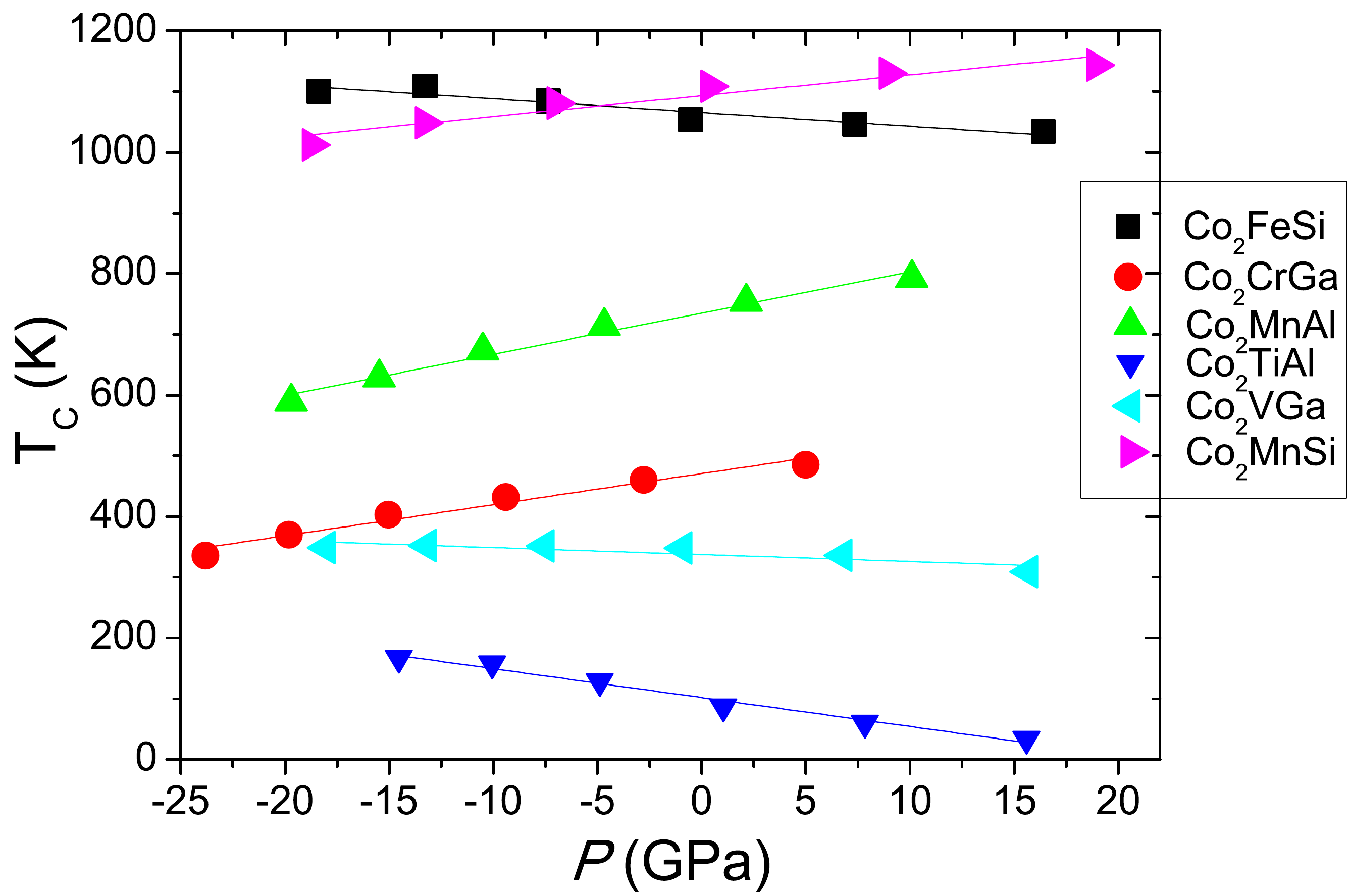}
\caption{Pressure dependence of the calculated Curie temperatures for all the Husler compounds studied. The pressure was found by fitting the total energies to the Birch-Murnaghan equation of state. The lines are linear fits.}
\label{fig_TCP}
\end{figure}

The results obtained at the experimental lattice parameter, shown in the table, are very close to previous calculations in similar conditions (within $2.5\%$ for moments and $6\%$ for $T_C$ at most, usually less)~\cite{Thoene09}, keeping the same good agreement with experimental values. 

Fig.~\ref{fig_TCs} shows the calculated Curie temperature and its coupling to the volume.  A variety of behaviors are found in these compounds. For Co$_2$FeSi, Co$_2$TiAl and Co$_2$VGa the temperature increases with volume, while it decreases for Co$_2$CrGa, Co$_2$MnGa, and Co$_2$MnSi. The variation is linear for the latter, while some non-linearities are observed in the former. Co$_2$FeSi, with detail in the lower part of the figure, even has a non-monotonic dependence, since $T_C$ reaches a maximum at the experimental parameter and decreases slightly when $a$ is increased 1\%. A nonmonotonic pressure dependence was also previously found for the case of Ni$_2$MnSn, when extending calculations for  higher positive pressures than measured. In our case, the extended lattice parameter of Co$_2$FeSi corresponds to a negative pressure, $-18$\ GPa. The spin moment, however, is still increasing. Therefore, it is only at this volume expansion that the hopping decrease (with increasing volume) starts to dominate over the moment increase, decreasing the overall exchange interactions. 

Comparing Fig.~\ref{fig_TCs} with Fig.~\ref{Dominant_J} for all the alloys, it is seen that the dominant exchange interaction already agrees qualitatively with the $T_C$ dependence, consistent with the  the fact the the only significant interactions should be for the first few neighbors.


%
%

%
\begin{table*}[ht!]
\renewcommand{\arraystretch}{1.1}
\caption{Curie temperatures and spin moments at the experimental lattice parameters, theoretical lattice parameter, bulk modulus, $dJ_{max}/da$, $dT_C/dV$, and $dT_C/dp$ near zero pressure.}
\label{table_exp}
\centering
\begin{tabular}{|l||c|c|c|c|c|c|c|c|}
\hline
Compound & $T_C^{\text{MFA}}$ (K)& $M_{tot}^{calc}$ ($\mu_B$) & $a^{exp.}$ (\AA{})& $a^{calc.}$ (\AA{}) & $B_P$ (GPa) & $dJ_{max}/da$ (meV/\AA{}) & $dT_C/dV$ (K/\AA{}$^3$) & $dT_C/dP$ (K/GPa)\\
\hline
Co$_2$TiAl & $157$ 
& $0.87$ & $5.84$ & $5.73$ & $202.9$ & $ 9.1 $ & $4.7$ & $-3.9$ \\
\hline
Co$_2$VGa & $352$ 
& $1.95$ & $5.78$ & $5.66$ & $237.0$ & $ -1.5 $ & $2.1$ & $-1.4$ \\
\hline
Co$_2$CrGa & $370$ 
& $2.98$ & $5.81$ & $5.61$ & $245.2$ & $ -5.3 $ & $-4.5$ & $3.2$ \\
\hline
Co$_2$MnAl & $629$ 
& $4.00$ & $5.75$ & $5.59$ & $228.0$ & $ -10.4 $ & $-7.4$ & $5.9$ \\
\hline
Co$_2$MnSi & $1049$ 
& $4.94$ & $5.65$ & $5.54$ & $258.0$ & $ -14.5 $ & $-2.4$ & $3.7$ \\
\hline
Co$_2$FeSi & $1109$ 
& $ 5.30 $ & $ 5.64 $ & $ 5.52 $ & $ 241.9 $ & $ 10.8 $ & $1.4$ & $-0.9$  \\
\hline
\end{tabular}
\end{table*}

An estimate of $dT_C/dP$ near zero pressure gives $-0.9$\ K/GPa for Co$_2$FeSi,  $-3.9$\ K/GPa for Co$_2$TiAl, $-1.4$\ K/GPa for Co$_2$VGa, $3.2$\ K/GPa for Co$_2$CrGa, $5.9$\ K/GPa for Co$_2$MnAl, and $3.7$\ K/GPa for Co$_2$MnSi. It could be expected that, at large interatomic distances, the effect of increasing hopping dominates against the effect of decreasing moments, producing a positive $dT_C/dP$~\cite{Sasioglu05}. 
Starting from smaller lattice constants (last rows of the table) the first compound has (mostly) a negative $dT_C/dP$ in the pressure range studied, while the next two compounds have a positive $dT_C/dP$,  increasing with lattice constant, which seems to be in agreement with this trend. However, the continued increase of the lattice constant, with the other three compounds, decreases $dT_C/dP$ and Co$_2$VGa and Co$_2$TiAl even have a negative $dT_C/dP$.  In fact it is not surprising that this trend, previously found for Mn compounds, is not followed for three of these compounds with different transition metals (Cr, V, Ti). In the last two cases, as previously noted,  the dominant pairwise $J$ is between Co1 and Co2, not between Co and the other transition metal, which is related to the negative sign of $dT_C/dP$, since that interaction (and the total spin moment) follows the same decreasing trend with pressure.  Other derivatives are presented in the table, describing the variation of the highest pairwise $J$ with $a$, and the variation of $T_C$ with volume.

The results shown in table~\ref{table_exp} can be used to calculate thermodynamic quantities, particularly the field induced magnetic entropy change (magnetocaloric effect). The magnetic moment, $T_C$ and $dT_C/dV$ can be employed in a mean-field model approach such as the Bean-Rodbell model~\cite{Bean_1962,Amaral_DICNMA2013}. This simple model allows the quick evaluation of the predicted order of the transition (second or first), and the contribution of the magnetovolume coupling to the total observed magnetic entropy change. Nevertheless, this approach ignores microscopic effects, which may be relevant for a more detailed analysis or material optimization. Alternatively, the value of $J$ and $dJ/da$ can be used for Monte-Carlo simulations of compressible latices, in the simplest case following the Domb model \cite{Domb_JChemPhys_1956}. This approach is more challenging, as the thermodynamic properties of these compressible lattice models remain largely unexplored.
 
\section{Conclusion}

In summary, we have calculated the effect of volume changes in the magnetic interactions in Co$_2$XY Heusler alloys. The results here obtained are a starting point to the use of either mean-field or microscopic models to estimate the magnetocaloric effect of a given material, as a theoretical approach to material optimization and the search for new high-performance magnetic refrigerant materials. The work here presented shows important tendencies to study and optimize with more extensive calculations. The use of more accurate exchange-correlation functional approximations may be required to obtain better estimates. Additionally, further optimization of $dT_C/dP$ can be done by exploration of fractional alloys via the use of the coherent potential approximation.

\section*{Acknowledgments}
This work is funded by FCT and FEDER-COMPETE, FCT grants SFRH/BPD/82059/2011 (J.~N. Gon\c{c}alves), SFRH/\-BPD/\-63942/\-2009 (J.~S. Amaral), and by projects PTDC/CTM-NAN/115125/2009, EXPL/\-CTM-NAN/\-1614/\-2013 (FCOMP-01-0124-FEDER-041688) and PEst-C/\-CTM/\-LA0011/\-2013 (FCOMP-01-0124-FEDER-037271).


\ifCLASSOPTIONcaptionsoff
  \newpage
\fi



%



\end{document}